\documentclass[a4paper, 11pt]{article}
\usepackage{jheppub}
\usepackage{graphicx}
\usepackage{tikz}    
\usepackage{animate}
\usepackage{color}        
\usepackage{pgfplots}     
\usepackage{xcolor}     
\newcommand{\be}{\begin{equation}}
\newcommand{\ee}{\end{equation}}
\newcommand{\ba}{\begin{eqnarray}}
\newcommand{\ea}{\end{eqnarray}}
\newcommand{\bi}{\begin{itemize}}
\newcommand{\ei}{\end{itemize}}
\newcommand{\tr}{{\rm Tr\,}}

\newcommand{\la}{\label}

\newcommand{\non}{{\nonumber}}

\def\a{\alpha}               
\def\d{\delta}        
          \def\l{\lambda}     \def\L{\Lambda}
\def\m{\mu}                     
\def\n{\nu}             
\def\r{\varrho}       
\def\t{\tau}          
              
        \def\z{\zeta}
  \def\OO{{\cal O}}

\newcommand\Tr{\mbox{Tr}}

\title{Caloron correction to the effective potential in thermal  gluodynamics} 
\author[a,b]{Chris P. Korthals Altes}
\author[c]{and Alfonso Sastre}

\affiliation[a]{Centre Physique Th\'eorique,
UMR 6207 of CNRS, Universities  Aix-Marseille I and II, 
and of Sud Toulon-Var, and affiliated with FRUMAM.
Case 907, Campus de Luminy, F-13288 Marseille, France, }
\affiliation[b]{NIKHEF theory group, P.O. Box 41882, 1009 DB Amsterdam, The Netherlands}   
\affiliation[c]{Bergische Universität Wuppertal - Fachgruppe Physik, Gaussstr. 20, 42119 Wuppertal, Germany}
  
\emailAdd{altes@cpt.univ-mrs.fr}
\emailAdd{sastrebruno@uni-wuppertal.de}
        
\abstract{
The one loop effective potential in thermal gluodynamics has stable minima in perturbation theory, where the Wilson line is  center group valued. This stays true 
to all loop orders.
However, calorons with non-trivial holonomy contribute to one loop order a linear term in the holonomy. This term is computed for the gauge group $SU(2)$.
The sign is such that the center group minimum stays stable.
}
 
\preprint{CPT-P003-2014, NIKHEF/2014-049, WUB/14-16}

\begin{document}

\maketitle

\section{Introduction}

The existence of a plasma phase of QCD has been confirmed by experiments at RHIC and LHC. 
Lattice simulations have been since long precious information about the equilibrium 
properties of the plasma.

One question of interest is how the Stefan-Boltzmann gas of quarks and gluons at very high
temperature develops into the strongly coupled plasma at lower temperatures, still above 
the critical temperature $T_c$.  

One can make the question more precise by looking at the free energy of the plasma as function
of the order parameter, the Polyakov loop (Wilson line). This order parameter measures the surplus
free energy of a heavy fundamental quark and reads for $SU(2)$
\be
\frac{1}{2}\Tr\mathcal{P}\exp\left(i\int_0^{1/T}dt A_0(\vec x,t)\right)=\frac{1}{2}\Tr P(A_0(\vec x))
\la{polyakovloop}
\ee
For a constant background $A_0=2\pi q T\frac{\t_3}{2} $ and the  Wilson line  is parametrized by
\be
\frac{1}{2}\Tr P(A_0(\vec x))=\cos(\pi q).
\la{polyakovloop1}
\ee

For the sake of discussion  we will exclude quarks from the system and do the calculation for 
gauge group $SU(2)$. In that case  the free energy is $Z(2)$ symmetric, i.e. periodic {\bf  mod 1} and invariant under exchange of $q$ 
and $1-q$. This symmetry is spontaneously broken at high temperature $T$, where the order parameter
does not vanish. The one loop fluctuations around the translation invariant state 
Eq. (\ref{polyakovloop1})  generate the free energy 
\be
f_{tr}=f_{SB}+\frac{4\pi^2}{3}T^4q^2(1-|q|)^2.
\la{translationfree}
\ee
 
The minima of this free energy represent the actual equation of state of the plasma. At very high
temperature this is the Stefan-Boltzmann (SB) law with the Polyakov loop being $\pm 1$.   At temperatures just above the critical 
temperature the minima will become nearly degenerate (first order transition) or coalesce at 
$q=\frac{1}{2}$ (second order transition, as indeed realized in $SU(2)$, and where the Polyakov loop
(\ref{polyakovloop1})  vanishes.). 

Perturbative corrections to the potential do not change the location of the minima as is expected.
One may ask:  does the location change when semi-classical  
 saddle points like those of thermal instantons or calorons are taken into account?  Do these saddle points initiate the
 movement of the SB minima?

 To obtain the  answer one has to compute the determinant of fluctuations around the saddle points in question. 
Its contribution to the free energy can be written on general grounds in terms of the running coupling $g(T)$ as:
\ba
f_{tot}&=&f_{tr}+f_{cal}\non\\
f_{cal}&=&-2T^4\left(\frac{4\pi^2}{g^2(T)}\right)^4\exp\left(-\frac{8\pi^2}{g^2(T)}\right)\Bigg(n_0+n_1 q+\cdots \Bigg)
\la{calcontrifree}
\ea
 
 The last factor in parenthesis is the linearized instanton density integrated over  the caloron size. 
 The dots are terms that render $f_{cal}$ $Z(2)$ symmetric.
 
 To leading order in the coupling $n_{0,1}$ are just {\it numerical constants}.  
 They pick up a temperature dependence  only in next to leading order, which warrants the calculation of two loop effects.

  The coefficient $n_0$ was calculated in the seminal paper by Gross, Pisarski and Yaffe
 ~\cite{Gross:1980br} (GPY in the sequel). It increases  the Stefan Boltzmann pressure.  The coefficient $n_1$ is computed in this paper.
 It has a negative  sign, such that the free energy $f_{cal}$ increases. Therefore  the minimum of the SB ground state
 at $q=0$ stays stable.

 To further motivate the work done below we mention the work Diakonov and coworkers~\cite{Diakonov:2004jn,Diakonov:2005qa,Gromov:2005ij,Gromov:2005hv}
 (from now on DGPS).
 They have computed the caloron determinant  as an expansion 
 in the inverse size $r_{12}$ of the caloron. So their result is most reliable at large size $r_{12}$. Now the coefficients $n_{0,1}$
 are  both obtained by an integral over the size. This integral converges in the ultra-violet due to asymptotic freedom (suppressing
 the small $r_{12}$ contribution by a power), and in the infrared due to thermal screening. The latter suppresses exponentially the large
 size contribution, where the expansion of DGPS is most precise.

   In this paper we avoid the large size expansion and obtain an analytic result for the linearized  density in Section \ref{sec:oneloop}, Eq. (\ref{calorondensity1}). 
   
  As shown below  this coefficient  can be  expressed  in terms of moments  of the GPY result for the instanton 
  density  at $q=0$. Recently the GPY result has been recast in analytic form~\cite{Altes:2014bwa}, so $n_1$ is a very well controlled quantity.

 Yet another  motivation for computing only the linear term $n_1 q$ comes from the dependence of the screening on the holonomy $q$. 
 It is known since long \cite{Giovannangeli:2002uv,Hidaka:2009hs} that 
 the screening mass in a constant background $\pi q T\t_3$ calculated for diagonal fields is proportional to the second derivative of $f_{tr}$
 \be
 \frac{1}{6}-q(1-q).
 \ee
 As screening is effective at large distance where the caloron by definition equals the holonomy $q$, and  the caloron fields are diagonal at large distance
 one would expect the screening  to be of this type. This is indeed what DGPS do find for the screening of the caloron.

\begin{figure}            
 \centering 
  \includegraphics[scale=1.00]{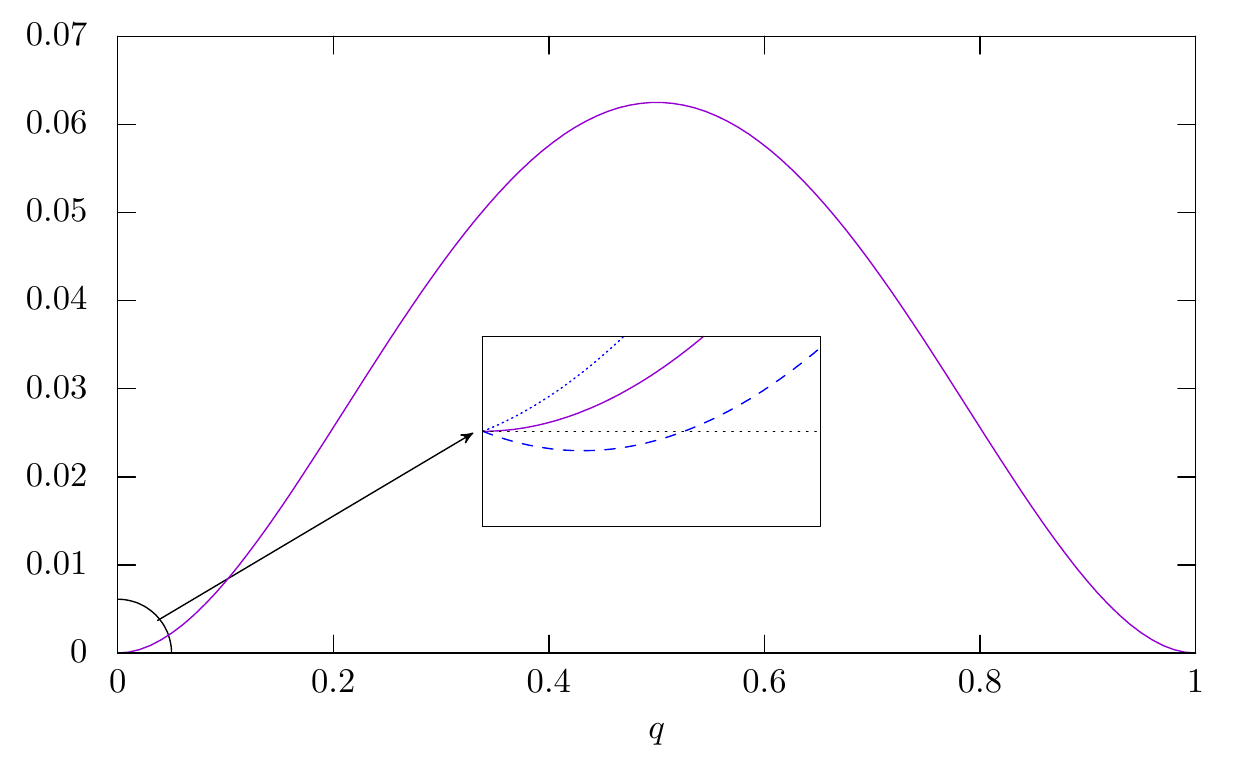}
  \label{fig:potential}
\caption{The potential (\ref{translationfree})and in the blow up  the  correction due to (\ref{calcontrifree}) with a  positive (dashed) and negative (dotted) linear term $q n_1$. 
The dotted line is realized by the caloron gas as shown in this paper. }
\end{figure}

 This screening turns into anti-screening in a window centered around $q=\frac{1}{2}$, where the potential is concave (see Fig. \ref{fig:potential}).
 Hence the contribution of the caloron will diverge in that window\footnote{A feature common in mean field approximations.}. 
 So one may limit oneself just as well to only the linear term, which by itself gives important information on the physics.

Knowledge of that term  determines the stability of the Stefan-Boltzmann minima. This is the strategy we will follow.

 Recapitulating, on the basis of our analytic result for this linear term we conclude that the minima do not change due to caloron fluctuations,
 they stay stable. Our statement is true to {\it leading} order in the running coupling $g(T)$.

The lay-out of the paper is as follows. In Section \ref{sec: sblowert} the effective potential is introduced,
and the main motivation for the paper explained. In particular the relevance of the term linear in a holonomy expansion is emphasized.
In Section \ref{sec:caloronclassics} the classical caloron solution is presented together with  a discussion of the variation
of the caloron with respect to the holonomy.  Appendix \ref{app:A} does furnish details. 
 
Section \ref{sec:oneloop} contains the central part of our paper.  
 In Section \ref{sec:results} our results are summarized and we compare them to those of DGPS, as far as possible.

\section{Effective potential and its behavior away from the Stefan- Boltzmann limit}{\la{sec: sblowert}}

In this section we discuss the effective potential to one loop order in a constant Polyakov loop background $\cos(\pi q)$.
More precisely the effective potential is defined as a Euclidean path integral with the gauge invariant constraint that only
configurations with 

\be
\frac{1}{2V}\int ~d\vec x Tr P(A_0)(\vec x)=\cos(\pi q)
\la{holonomyconstraint}
\ee
\noindent are admitted~\cite{KorthalsAltes:1993ca}
where obviously $q$ is constant in space.  The three volume $V$ is supposed to be macroscopically large.   
An obvious choice for the vector potential is
 \be
 A_\m=\d_{\m,0}2\pi T q\frac{\t_3}{2}.
 \la{constantbackground}
 \ee  
The classical action vanishes for all $q$, 	 so is degenerate. 
The fluctuations  around this translation invariant background do lift this degeneracy.   
  
The result to one loop \cite{Gross:1980br,Weiss:1980rj} and two loop \cite{Bhattacharya:1992qb,Dumitru:2013xna,Guo:2014zra}
 is a very simply  $q$-independent renormalization    
of the one loop result 
\be
f_{tr}=\Bigg(f_{SB}+\frac{4\pi^2}{3}T^4q^2(1-|q|)^2\Bigg)\left(1-\frac{10g^2}{(4\pi)^2}\right).
\la{translationinvfree}
\ee

The $\OO(g^2)$ term is due to the two loop free energy diagrams in the background $q$
\footnote{And a renormalized Polyakov loop insertion into the one loop free energy, 
which renders the result  (\ref{calcontrifree}) gauge independent and kills all linear terms in $q$ }.
The Stefan-Boltzmann free energy is
\be
f_{SB}=-\frac{\pi^2}{15}T^4.
\ee

Clearly the minima are in $q=0,1$ where the loop takes the $Z(2) $ values  $\pm 1$, 
and this is still true to two loop order\footnote{This simple multiplicative renormalization is true for any group, 
with the factor 10 replaced by $5C_2$, the adjoint second order Casimir invariant\cite{Dumitru:2013xna}}.

The question is now:  can we find other local minima of the action with fluctuations  
furnishing for small holonomy  a {\it linear} term in $q$? If we choose the caloron minimum  then $f_{tr}$ changes into:
\be
f_{tr}+f_{cal}+\cdots
\la{linearterminftr}
\ee
\noindent with $f_{cal}$ as in Eq. (\ref{calcontrifree}).

 Then a negative coefficient  $n_1$ will destabilize the minimum at $q=0$. 
 A positive sign stabilizes the minimum. See Fig. \ref{fig:potential} and the blow up.
 
Local minima of that kind are provided by (anti)-self-dual configurations with topological charge $\pm 1$, 
and as boundary condition at spatial infinity that the 
Polyakov loop is $\cos(\pi q)$. Such configurations have classical action $8\pi^2/g^2$ 
, independent of the value of $q$.
However one loop fluctuations lift the degeneracy, and indeed provide us with a linear term. 
However we find the sign to be positive and hence the minimum stays where it is: at $q=0$.

\section{The caloron with non-trivial holonomy and the Harrington-Shephard thermal instanton}\la{sec:caloronclassics}

The caloron is an (anti)-self-dual solution of the Euclidean equations of motion, with the boundary condition that 
at spatial infinity the Wilson line $P(A_0)(\vec x)$ takes the value:
\be
\frac{1}{2}\Tr P(A_0)(\vec x)=\cos(\pi q).
\ee
In the trivial holonomy case ($q = 0$), the caloron gauge field in the singular gauge is given by the Harrington-Shepard (HS) solution
\cite{Harrington:1978ve}:     
\begin{eqnarray}
 A_\mu &=& -\bar\eta^a_{\mu\nu}\partial_\mu \log \Pi \\
\Pi &=& 1 + r_{12} f \\
f &=& \frac{1}{r}\frac{\sinh(r )}{\cosh(r)- \cos(t)}
\la{HSpot}
 \end{eqnarray}
where for convenience $r$ and $t$ have been rescaled $r \rightarrow 2\pi r T$ and $t \rightarrow 2\pi t T$
\footnote{The partial derivative will always be  in the unscaled coordinates}.
The 't Hooft symbol  is given by:
\be
\sigma_{\mu}\sigma_\nu^\dagger = \delta_{\mu\nu} + i\bar\eta_{\mu\nu}^a \tau^a
\ee
where $\sigma_\mu = (1,-i\vec\tau)$, $\vec \t$ the Pauli matrices. 

In the general case ($q\not=0$), the caloron can be seen as composite by two constituent monopoles 
\cite{Kraan:1998kp,Lee:1998bb,Kraan:1998pm}. At large separation $r_{12}$ one is approximately a 
BPS \cite{Bogomolny:1975de,Prasad:1975kr} monopole, the other is a twisted  BPS monopole (in short 
a Kaluza-Klein (KK) monopole \cite{Rossi:1978qe}). 
They have opposite magnetic charges and topological charges $q$ and $1-q$ respectively.
So the solution has topological charge $1$ and vanishing total magnetic charge. 

The action equals $8\pi^2/g^2$ and is degenerate in the zero mode parameters and the holonomy $q$.

Using rotational and translation symmetry we can put the two monopoles on  the $z$ axis
and the centre of mass at the origin.
The separation between the constituent monopoles is fixed by $r_{12}$:
\begin{eqnarray}
 q \vec X_1 + (1-q) \vec X_2 &=& 0\\
  \vec X_2 - \vec X_1 &=& (0,0,r_{12})
\end{eqnarray}
The sizes of the BPS and KK monopoles are determined by the inverse of $q$ and $1-q$ respectively, so we have two different
scenarios depending on  whether $q^{-1} \ll r_{12}$ or $q^{-1}\gg r_{12}$.   
In Figures \ref{fig:smallcal} and \ref{fig:largecal} we present an example of each case:   
\begin{itemize}
 \item  Figure \ref{fig:smallcal} corresponds to $q r_{12} \gg 1$: in this case both
 constituent monopoles are well separated. This is the case analysed in \cite{Diakonov:2004jn,Diakonov:2005qa,Gromov:2005ij,Gromov:2005hv}
 with a large $r_{12}$ expansion.
\begin{figure}
\centering  
\begin{tikzpicture}   
\pgfmathsetmacro{\mdelta}{0.2}
\pgfmathsetmacro{\mrho}{1.} 
\pgfmathsetmacro{\mypi}{3.141519}
\pgfmathsetmacro{\mone}{4*\mypi *\mdelta}
\pgfmathsetmacro{\mtwo}{4*\mypi *(0.5-\mdelta)}
\pgfmathsetmacro{\Xone}{(-1+2*\mdelta)*\mrho *\mrho *\mypi}
\pgfmathsetmacro{\Xtwo}{ 2*\mdelta *\mrho *\mrho *\mypi }

\draw[fill=red] (\Xone,0) circle (0.1) node[below] {$X_{1}$} ;
\draw[dashed,fill=red,opacity=0.1] (\Xone,0) circle (1/\mone) ;
\draw[dashed,->] (\Xone,0) -> node[left] {$q^{-1}$} (\Xone,1/\mone) ;

\draw[fill=blue] (\Xtwo,0) circle (0.1) node[below] {$X_{2}$} ;
\draw[dashed,fill=blue, opacity=0.2] (\Xtwo,0) circle (1/\mtwo) ;
\draw[dashed,->] (\Xtwo,0) -> node[right] {$\bar q^{-1}$} (\Xtwo,1/\mtwo) ;

\draw[fill=black] (0,0) circle (0.05) node[below] {$X_{CM}$} ;
\draw[dashed] (\Xone,0) --  (\Xtwo-0.1,0);  
\draw[<->] (\Xone,-0.5) -- node[below] {$r_{12}$} (\Xtwo-0.1,-0.5);  

\draw[dashed,->] (\Xone ,0) -- node[below,left] {$r_1$} (1.8,2.0);  
\draw[dashed,->] (\Xtwo  ,0) -- node[below,right] {$r_2$} (1.8,2.0);  
\draw[dashed,->] (0,0) --  (1.8,2.0) node[right] {$r$};  
\end{tikzpicture}
\caption{Caloron with size $r_{12}$ larger than  the charge radii $q^{-1}$ and ${\bar q}^{-1}$ of HS and KK monopoles}  
 \label{fig:smallcal}
\end{figure}
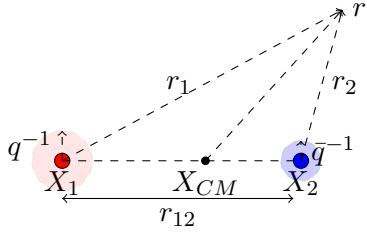

 \item  Figure \ref{fig:largecal} corresponds to $q r_{12}\ll 1$: 
 in this case the KK monopole is embedded in the core of  the  BPS monopole. It is  instructive to see how the rotational symmetry is still broken through terms of $\OO(q)$, i.e. as long as the core of the BPS monopole is finite, but large. Only when $q=0$ the core is delocalized, and rotational invariance is exact.  This physics is better reflected in a parametrization \cite{GarciaPerez:2006rt,GarciaPerez:2009mg} who use prepotentials that are specifically connected to the amount of breaking of rotational invariance. It reads
 \begin{eqnarray}
 A_\mu = -\bar{\eta}^a_{\mu\nu}\hat F\mathcal D_\nu\left(\delta_{ab}
 V_{S} 
 +\epsilon_{a3b}
   V_{A} 
 +\delta_{3a}\delta_{3b}
  V_{3}\right)\frac{\tau^b}{2}
 \la{BCEbasis0}
\end{eqnarray}
 
This expression is quite useful since $\mathcal{D_\n}$ is a covariant derivative common to the  scalar potentials $V_{S,A,3}$.
Details are to found in Appendix \ref{app:A}, where also  the relation to the parametrization of KvB is given.  
Note that the tensor $\bar\eta_{\m\n}$ is $SO(3)$ invariant, the other two tensors  {\it only} $SO(2)$ invariant.
Terms linear in $q$ in  the effective potential will be entirely controlled  by the potential with the largest symmetry,
$V_{S}$.  The  reason is that  the trace 
 in the loop calculation  obliterates the terms with  lower symmetry in $V_S$ and $V_3$. 
 An accidental symmetry cause $V_3$ to have vanishing  $\OO(q)$ terms.

\begin{figure}
\centering
\begin{tikzpicture}
\pgfmathsetmacro{\mdelta}{0.02}
\pgfmathsetmacro{\mrho}{0.8} 
\pgfmathsetmacro{\mypi}{3.141519}
\pgfmathsetmacro{\mone}{4*\mypi *\mdelta}
\pgfmathsetmacro{\mtwo}{4*\mypi *(0.5-\mdelta)}
\pgfmathsetmacro{\Xone}{(-1+2*\mdelta)*\mrho *\mrho *\mypi}
\pgfmathsetmacro{\Xtwo}{ 2*\mdelta *\mrho *\mrho *\mypi }

\draw[fill=red] (\Xone,0) circle (0.1) node[left] {$X_{1}$} ;
\draw[dashed,fill=red,opacity=0.1] (\Xone,0) circle (1/\mone) ;
\draw[dashed,->] (\Xone,0) -> node[left] {$q^{-1}$} (\Xone,1/\mone) ;

\draw[fill=blue] (\Xtwo,0) circle (0.1) node[right] {$X_{2}$} ;
\draw[dashed,fill=blue, opacity=0.2] (\Xtwo,0) circle (1/\mtwo) ;
\draw[dashed,->] (\Xtwo,0) -> node[above] {$\bar q^{-1}$} (\Xtwo,1/\mtwo) ;

\draw[fill=black] (0,0) circle (0.05) node[below] {$X_{CM}$} ;
\draw[dashed] (\Xone,0) --  (\Xtwo-0.1,0);  
\draw[<->] (\Xone,-0.5) -- node[below] {$r_{12}$} (\Xtwo-0.1,-0.5);  
   
\draw[dashed,->] (\Xone ,0) -- node[below,left] {$r_1$} (1.8,2.0);  
\draw[dashed,->] (\Xtwo  ,0) -- node[below,right] {$r_2$} (1.8,2.0);  
\draw[dashed,->] (0,0) --  (1.8,2.0) node[right] {$r$};  
 \end{tikzpicture} 
\caption{Caloron with size $r_{12}$ smaller than the charge radius of the BPS monopole. } 
\label{fig:largecal}  
\end{figure}
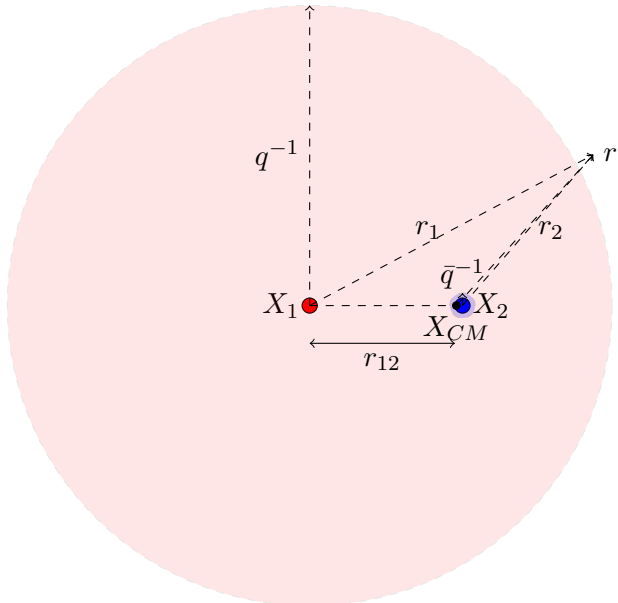

The result in the limit $q r_{12} \ll 1$ for the caloron gauge potential can be written as (see Appendix \ref{app:A}):
\begin{eqnarray}
 A_\mu^{a} &=&  -\bar\eta^a_{\mu\nu}\partial_\nu \left(\log \left(1+ r_{12}(1 +r_{12} q)f\right)\right)  
 + q(\mathrm{ NRI ~terms})+ \mathcal{O}(q^2) 
\label{eq:smallq} 
 \end{eqnarray}   
The first term in Eq. (\ref{eq:smallq}), corresponds to a HS caloron 
with size $r_{12}$ changed into $r_{12}(1 + q r_{12})$.  The terms that are not 3d rotational invariant are indicate by NRI.

This requires the computation of the variation of the caloron configuration with respect to $q$. 
Eq. (\ref{eq:smallq}) is the result of two  observations:
\begin{eqnarray}   
\partial_{r_{12}} A_\mu^a(r_{12},q=0) &=& -\frac{1}{r_{12}}\bar\eta^a_{\mu\nu} \frac{\partial_\nu \Pi}{\Pi^2}\\
\partial_q A_\mu^a(r_{12},q)|_{q=0} &=& -r_{12}\bar\eta^a_{\mu\nu} \frac{\partial_\nu \Pi}{\Pi^2} 
+ (\mathrm{NRI ~ terms})
\end{eqnarray}
The first relation is simply expressing the linearity of the HS caloron in terms of $r_{12}$, Eq. (\ref{HSpot}). The second relation is the result of the 
calculations in Appendix \ref{app:A}.

The two relations combine into
\begin{eqnarray}
\partial_q A_\mu^a(r_{12},q)|_{q=0} = r_{12}^2 \partial_{r_{12}} A_\mu^a(r_{12},q=0)  +
(\mathrm{NRI~ terms}),
\la{key}
\end{eqnarray}
\end{itemize} 
\noindent which is nothing but the term linear in $q$ in Eq. (\ref{eq:smallq}), using the HS potential Eq. (\ref{HSpot}).

This is our key result. 

 As we will see in Section \ref{sec:oneloop} the NRI terms are washed out in the one loop integration.
This means the linear term is determined, up to some minor modification (see Section \ref{sec:oneloop}) 
by the $r_{12}$ derivative of the HS determinant, something we had already computed in our previous paper \cite{Altes:2014bwa}.

\section{One loop contribution of the caloron}\la{sec:oneloop}

The one loop contribution is given by the fluctuation determinant, integrated over the zero modes. 
Once the zero modes are factored out the fluctuation determinant of the spin one particle (written with a prime below)
equals the square of that  of the spin zero particle. The latter has no zero modes. Taking the ghost into account we get:
\ba
Z_1&=&\int d\z J(\z) \left(\frac{4\pi^2}{g^2}\right)^{-4}\exp\left(-\frac{8\pi^2}{g^2}\right){\det}^{-1/2} (-D'_{\m\n})\det(-D^2_{gh})\non\\
&=&\int d\z J(\z){\det}^{-1} (-D^2). 
\la{transition}
\ea

The logarithmic divergencies are cured by normalizing  $Z_1$ with $Z_1^{(0)}$, the determinant of the zero temperature instanton\cite{'tHooft:1976fv}\footnote{This normalization is used in Eq. (\ref{eq:barA})}.  

This determinant in turn 
 contains the one loop short distance effects for the coupling
\be
\frac{8\pi^2}{g^2(x)}=\frac{22}{3}\log(x) +\log\log\mbox{ terms}.
\ee
With $\L$ the Pauli-Villars scale equal to $e^{\frac{1}{22}} \L_{\overline{MS}}$ one gets. 
\be
x=\frac{1}{\r\L}=\sqrt{\frac{2}{r_{12}}}\frac{\pi T}{\L}, ~r_{12}=2\pi^2\r^2 T^2.
\ee

\noindent  $\zeta$ stands for the eight zero modes parameters and $J(\zeta)$ is the zero mode measure
 \be
J(\z)=1+r_{12}q(1-q)
\la{zeromodemeasure}
\ee

\noindent The covariant isospin one d'Alembertian $D^2$ for a scalar particle reads:
\be
(\partial_\m\d_{ac}+iA_\m^d\epsilon_{adc})(\partial_\m\d_{cb}+iA_\m^e\epsilon_{ceb})\equiv D^2_{ab},
\la{isospinoneDalembertian}
\ee 
where the  potential  $A_\m$ is the caloron potential, Eqs. (\ref{kvb1}) and (\ref{kvb2}).  
The zero modes do not appear in the result of the determinant, except for the scale factor $\r$. 
The zero mode integration then  takes the form from Eq. (\ref{zeromodemeasure}):

\be
\int d\z J(\z)=\frac{V}{T} \n \int_0^\infty {d\r\over{\r^5}}(1+r_{12}q(1-q))=\left(\frac{V}{T}\right)
\n 2\pi^4T^4\int{dr_{12}\over{r_{12}^3}}(1+r_{12}q(1-q))
\la{zeromodemeasure1}
\ee

$\n$ is the volume of the $SU(2)$ instanton.  The space time volume 
 $V/T$ is due to the space time zero mode integration.

The transition amplitude  $Z_1$ is usually written as a product of the translation invariant  part, 
and a part due to the caloron 
\ba
Z_1&=&Z_{tr}\hat Z_1\\ \la{separationtranslatcaloron}
\hat Z_1&=&\left(\frac{V}{T}\right)T^4\left(\frac{4\pi^2}{g^2(T)}\right)^4\exp\left(-\frac{8\pi^2}{g^2(T)}\right)
\int_0^\infty dr_{12} n(r_{12},q, T/\L)
\la{calorondensitydefined}
\ea

Here $T^4 \left(\frac{4\pi^2}{g^2(T)}\right)^4\exp\left(-\frac{8\pi^2}{g^2(T)}\right)n(r_{12},q, T/\L)$ 
does represent the  density per unit space time volume of  calorons with size $r_{12}$,  temperature $T$ and holonomy $q$.
To leading order in the coupling there is {\it no} $T$ dependence in $n(r_{12},q, T/\L)$, only in the prefactors.
  
The one loop contribution of the translation invariant minima, i.e. with $A_\m=\d_{\m,0}2\pi qT \frac{\tau_3}{2}$ is $f_{tr}$, 
as given in Eq. (\ref{translationinvfree})  
\be
Z_{tr}=\exp(-{V/T} f_{tr})
\la{translationinvfree1}
\ee
 This bulk free energy  gets a correction due to the caloron gas:  
 \be
\sum_{N_\pm}{\hat Z_1^{N_+}\over{N_{-}!}}{\hat Z_{-1}^{N_-}\over{N_-!}}=\exp (2\hat Z_1)
\la{grand}
\ee
Combining Eqs. (\ref{separationtranslatcaloron}), (\ref{translationinvfree1}) and (\ref{grand})
one obtains for the total free energy 
\be
f_{tot}=f_{tr}-2\hat Z_1/(V/T).
\ee

So the free energy of the caloron becomes in terms of the density:
\ba
f_{cal}&=&-2 T^4\left(\frac{4\pi^2}{g^2(T/\L)}\right)^4\exp\left(-\frac{8\pi^2}{g^2(T/\L)}\right)\int_0^\infty dr_{12}n(r_{12}, q,T/\L),\non\\
n(r_{12}, q,T/\L)&=&C\int_0^\infty dr_{12}r_{12}^{2\over 3}\times (1+r_{12}q(1-q)) \exp(-\log\det(-D^2))\,.
\la{totalfreeisoone}
\ea    
The constant $C$ equals
\be
C=2^{-{8\over 3}} \n \pi^4.
\la{norm}
\ee

The  latter two equations tell us that the contribution to the bulk free energy equals minus twice the total density of calorons
as a function of $q$, as defined in Eq. \ref{separationtranslatcaloron}.  
From its definition in Eqs. \ref{transition} and \ref{zeromodemeasure1} (positive caloron measure) 
and Eq. \ref{separationtranslatcaloron} one expects this total density to be positive,
for all values of the holonomy $q$, for which it is defined. 

For vanishing holonomy one retrieves the density of Harrington Shepard calorons and the result of GPY\cite{Gross:1980br}.
The caloron gas produces a positive correction to the Stefan-Boltzmann pressure $-f_{SB}$.

\subsection{The linear term in caloron effective potential and 3d rotational invariance}
  
Clearly the caloron free energy, Eq. \ref{totalfreeisoone}, has  a linear term in $q$. First off all  it  receives
a contribution from the measure term $\sim r_{12}q(1-q)$, involving only the HS determinant. This contribution
is clearly negative, so destabilizes the minimum.   

That the  determinant    $ \exp(-\log\det(-D^2))$ delivers a linear term as well is shown by computing the derivative and setting $q=0$:
\ba
\partial_q\exp(-\log\det(-D^2))|_{q=0}&=&-\partial_q\left(\log\det (-D^2)\right) \exp(-\log\det(-D^2)|_{q=0})\non\\
&=&\Tr\left[(\partial_q D^2) (-D^2)^{-1}|_{q=0}\right]  \exp(-\log\det(-D^2)|_{q=0}).
\la{lineartermisoone}
\ea
The trace stands for space integration, integration over the time period and tracing of color.

The variation of the d'Alembertian at vanishing holonomy involves
\be
\partial_q D^2=\partial_q A_\m D_\m+D_\m\partial_q A_\m
\la{varA}
\ee 
The last two equations show that the covariant derivative $D_\m$, the propagator $(-D^2)^{-1}$ and the determinant are needed,
but {\it only} at vanishing holonomy, i.e. they are rotational invariant.

Note that the variation of the d'Alembertian on the right hand side of (\ref{lineartermisoone}) is projected on the rotational
invariant HS propagator at $q=0$\footnote{The HS propagator is only used as a regularization device, 
at coinciding points where it rotationally invariant.}.  In that projection only the rotational invariant part of $\partial_q A_\m$,  Eq. (\ref{key}), survives.  
In other words,  the linear term  in Eq. (\ref{lineartermisoone}) becomes, upon substitution of Eq. (\ref{key}) and using Eq. (\ref{varA}), 
\ba
\Tr\left[(\partial_q D^2) (-D^2)^{-1}\right]_{q=0}&=&r_{12}^2\Tr\left[(\partial_{r_{12}}A_\m) D_\m+D_\m(\partial_{r_{12}}A_\m)(-D^2)^{-1})\right]_{q=0}
+\Tr[ \mathrm{NRI}] \non\\
&=&r_{12}^2\Tr\left[(\partial_{r_{12}}D^2)(-D^2)^{-1}\right]_{q=0}+\Tr [\mathrm{NRI}].
\la{lineartermishs}  
\ea
The second equality follows because in the first line also the vector potentials $A_\m$ are zero holonomy  potentials.
The second set of terms vanishes for the mentioned symmetry reasons.  
The NRI terms are made explicit in Appendix \ref{app:A}, Eq. (\ref{resulttaylorexpansion}). 
It is then straightforward to check they do vanish  in the trace Eq. (\ref{lineartermisoone}).

So we conclude that the variation in $q$ at zero holonomy is just the variation in $r_{12}$ of the HS determinant, apart from an overall factor $r_{12}^2$.

This means that the trace involved in Eq. (\ref{lineartermisoone}), defining the coefficient of the linear term, is just
as convergent as the derivative of the HS determinant 
with respect to the size $r_{12}$\footnote{This convergence excludes terms like $q\log(q)$ in the effective action.}.

\subsection{The linearized instanton density}

To leading order the density  $n(r_{12},q,T/\L)$ is temperature independent.
The instanton density is  positive for any value of $\l$, and $q$,
\be
n(\l,q,T)=n_0(\l,q=0,T)+(n(\l,q,T)-n(\l,q=0,T))\ge 0.
\ee

In the preceding sections we have approximated the term in parenthesis by the linear term,
\be
n(\l,q,T)=n_0(\l,q=0,T)+q n_1(\l,T)+ \cdots
\la{lineardensity}
\ee 
\noindent and noticed (see Eq. (\ref{lineartermishs})) that we need only  the  HS  determinant~\cite{Altes:2014bwa}

\ba
\det( -D^{-2})|_{q=0}&=&\exp\left(-\left(16A+\frac{2}{3}r_{12}\right)\right)\non\\
 A&=&-{1\over{12}}\log\left(1+{r_{12}\over 6}\right)+R\non\\   
R&=&0.0129\left(1+0.899\left(\frac{r_{12}}{2}\right)^{-3/4}\right)^{-8}
\la{hsinput}     
\ea

$R$ is uniformly quite small compared to the logarithm.

The function $A$ is the result of a logarithmically diverging $\bar A$ subtracted with the one instanton  contribution $\bar A_0$. Explicitly:
\begin{eqnarray}
A(r_{12}) &=&{1\over{12}}{1\over {16\pi^2}}\left(\int_{R^3\times S^1} dx^4  \frac{\left((\partial_\mu \Pi)^2\right)^2}{\Pi^4}
-\int_{R^4} dx^4  \frac{\left((\partial_\mu \Pi_0)^2\right)^2}{\Pi_0^4}\right)\label{eq:barA}
\end{eqnarray}        
\noindent with $\Pi_0$ the one instanton prepotential:
\be
\Pi_0=1+{\r^2\over x^2}
\ee

Some care has to be taken in applying (\ref{lineartermisoone}) to the HS result. The reason is that $\bar A_0$ is not depending on the holonomy $q$. 
Hence it drops out when we vary the regulated  determinant  with respect to $q$, as in the first factor in (\ref{lineartermisoone}).

Hence using the input of  (\ref{hsinput})  the integrated caloron density $\hat Z_1$ becomes
\be
n_0+qn_1=C\int_0^{\infty} dr_{12}r_{12}^{2/3}\left(1+qr_{12}(1-r_{12}\partial_{r_{12}}\left(16 A+\frac{1}{3}\log(r_{12})+\frac{2}{3}r_{12}\right)\right)
e^{-\left(16A+{2\over 3}r_{12}\right)}.   
\la{calorondensity1}
\ee
 The term
 \be
 \frac{1}{3}\log(r_{12})
 \ee
 
 \noindent results from omitting the one instanton subtraction.

\section{Results and comparison to other work}
\label{sec:results}

\begin{figure}            
 \centering 
  \includegraphics{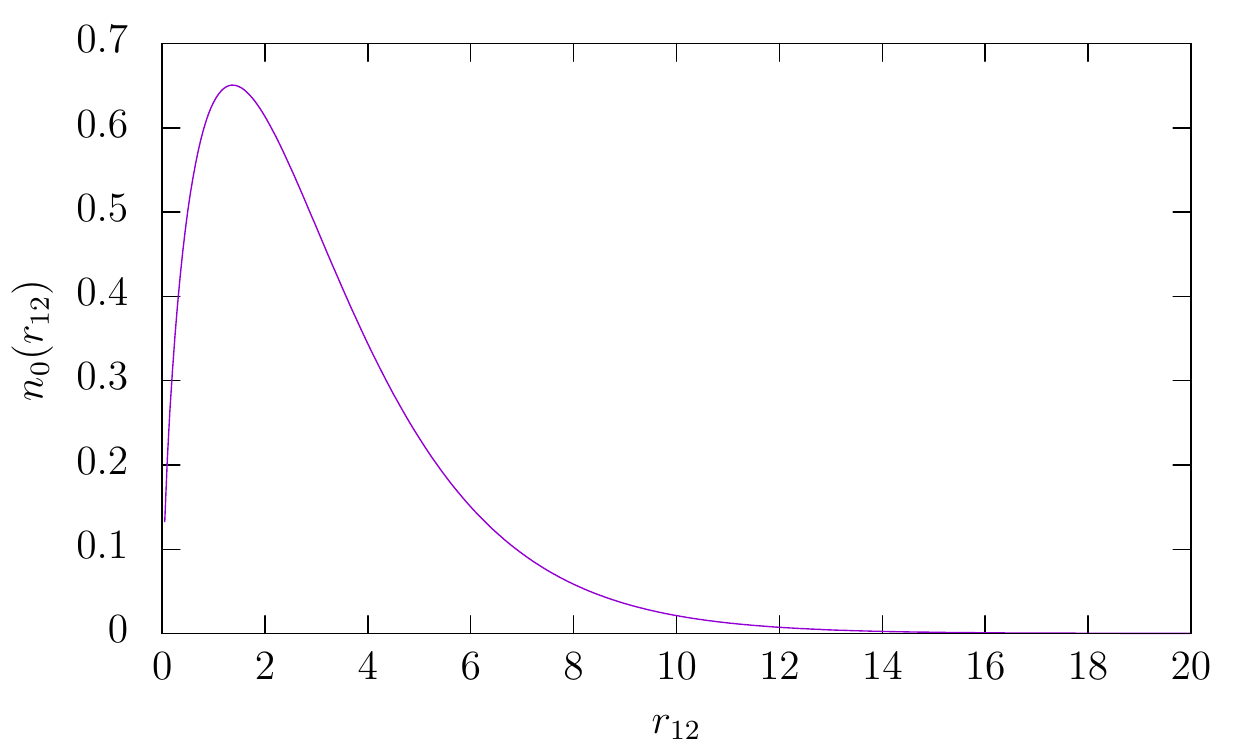}
  \includegraphics{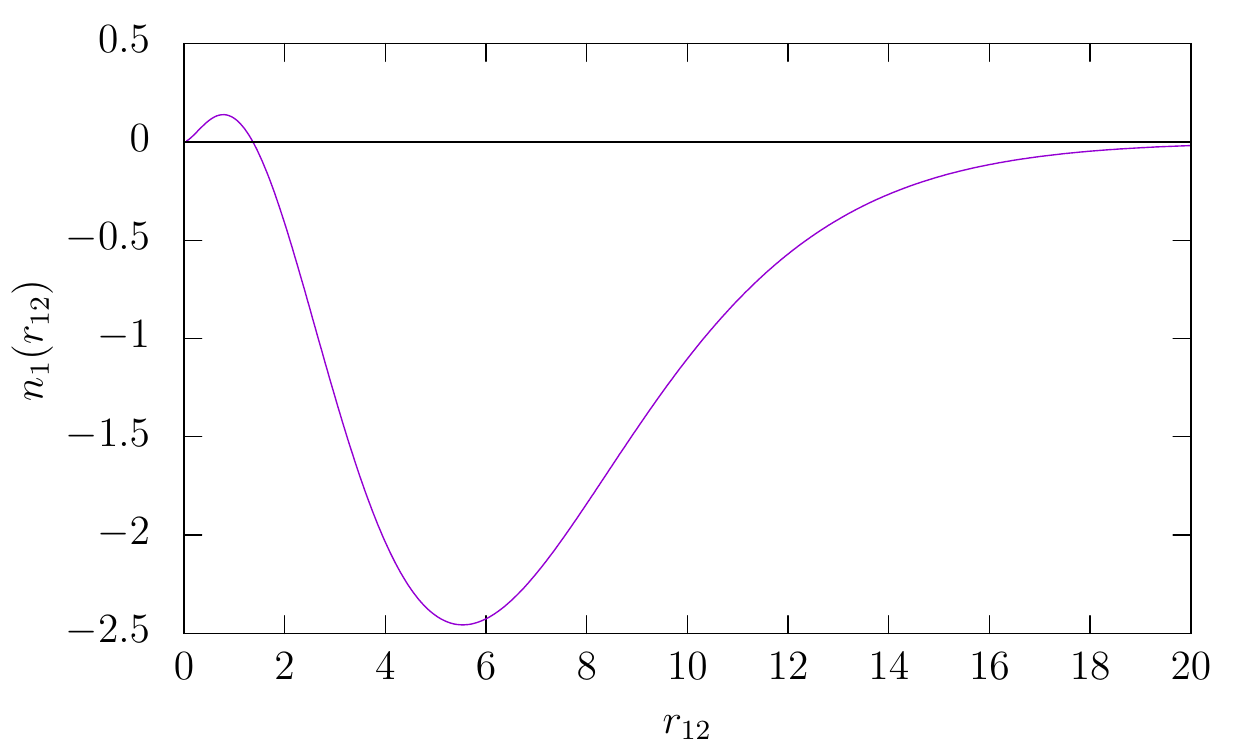}
\caption{Upper panel: the HS caloron density $n_0(r_{12})$ ;  
lower panel: the density of the coefficient $n_1(r_{12})$ \label{fig:density}}  
\end{figure}

Substituting Eq. (\ref{hsinput})
into Eq. (\ref{calorondensity1}) gives then the final result for the density $n_1(r_{12})$ and is shown in Fig. \ref{fig:density} 
together with the density $n_0(r_{12})$.

The outcome for the sign of $n_1$ is determined by a competition between the measure factor 
Eq. (\ref{zeromodemeasure1}) and the contribution from Eq. (\ref{totalfreeisoone}) as is clear from Fig. (\ref{fig:density}) or from inspection 
of the integrand multiplying $q$  in Eq. (\ref{calorondensity1}).   It is the latter that wins, with the result
\be
\int_0^\infty dr_{12} (n_0(r_{12})+q n_1(r_{12})=C(2.83854-17.5653q)  
\la{final}
\ee
Substituting this in Eq. (\ref{calcontrifree}) 
we get for the total free energy:       
\ba
f_{tot}&=&f_{tr}-2T^4\left({4\pi^2\over g^2(T)}\right)^4\exp\left(-{8\pi^2\over g^2(T)}\right)C\Bigg(2.83854-17.5653q+\cdots \Bigg)\,,\non\\
f_{tr}&=&f_{tr}=f_{SB}+{4\pi^2\over 3}T^4q^2(1-|q|)^2.
\la{finalfinal}
\ea
We see the  caloron determinant  gives  a positive correction to the pressure $-f_{SB}$ and a positive linear term to the  potential.
Clearly the latter does not change when varying the temperature. 

Let us compare this to the work of DGPS.
First in reference \cite{Gromov:2005ij} there is a $q\log(q)$ term in caloron contribution. Such a term would give a logarithmically diverging coefficient $n_1$.
However we have seen that that coefficient is just as convergent as $n_0$. 
On the other hand
 DGPS (see the relevant figure in  \cite{Gromov:2005ij})  do find from their numerical  approach that below a certain temperature the caloron fluctuations do indeed destabilize the 
 perturbative minima, but not above. That is, their first derivative of the effective potential  at the origin is positive.
 So in that sense their result agrees with ours at very high temperature.

 Since  the slope  in Eq.~(\ref{finalfinal})  is positive for all values of $T$, the DGPS result that  the slope at lower  $T$ switches sign must involve non-leading $\OO(g^2)$
 effects.
 
  The non-leading terms are partly due to
 two loop renormalization effects of the couplings. This is what is taken into account in the DGPS approach,
 are easy to compute and give indeed a negative $\OO(g^2)$ term. 
 and may indeed lead to a flip in the sign of Eq.~(\ref{calcontrifree}) if the coupling is large enough. 
 However the  two loop contribution (as yet to be calculated) may change that conclusion.

The nice observation of DGPS that the screening term in $\log\det(-D^2)$ equals
\be
4r_{12}\left({1\over 6}-q(1-q)\right)
\la{screening}
\ee
is strictly valid when the two monopoles are well separated. 
To illustrate this remember our linear term gets its contribution from the region where the monopole 
with core $1/q$ is embedding the KK monopole, the one that survives in the HS limit.  This is shown in Fig. \ref{fig:largecal}. 
In this regime we found, Eq. (\ref{eq:smallq}), that including the linear term just changes the size of the HS caloron:
\be
r_{12}\rightarrow r_{12}(1+qr_{12}).
\ee
So starting from the HS screening ${2\over 3}r_{12}$ we find a positive linear term ${2\over 3}qr_{12}^2$. That means switching on $q$ {\it increases} the screening!  
This is precisely opposite to what the screening Eq. (\ref{screening}) tells us. This is another illustration of the difference between the 
two regimes, well separated monopoles and one  monopole nested inside  the other. DGPS devote an interesting effort to the connection between  the two regimes.

Finally we offer  an amusing speculation. The potential increases from the SB minimum due to the caloron gas. 
However for larger $q$ the potential may bend and go down creating a stable minimum at some temperature $T_{ms}$, so that the SB minimum becomes meta-stable.  The  tunneling to this new minimum would give a glitch in the thermodynamic bulk functions.

What might be the moral  that  switching on the caloron gas  fails to initiate  the instabilty of the Stefan-Boltzmann gas ? A cynic would say, that calorons are after all  instantons and the latter are known since long
to fail to give a first step to confinement.  Proponents would say, that the monopoles inside the instanton might have  given precisely that first step.  Our computation shows that, alas, monopoles are not sufficient for destabilization.

\section{ Acknowledgements}
 Both authors thank the Centre de Physique Th\'eorique for its hospitality during the beginning of this work.
CPKA is indebted to NIKHEF  and BNL for hospitality. AS  
is funded by the DFG grant SFB/TR55 and  both authors thank the Wuppertal theory group for its hospitality.
We have enjoyed useful discussions with Rob Pisarski, Zoltan Fodor and Szabolc Borsanyi.

\appendix
\section{KvBLL caloron and its Taylor expansion around the HS caloron} \la{app:A}
In this appendix we give some details about the determination of the Taylor expansion 
of the KvBLL caloron around the HS caloron in terms of the holonomy, $q$. 
In order to simplify this calculation, we found the notation introduced in \cite{GarciaPerez:2006rt,GarciaPerez:2009mg} very
useful. 
Of course we compare with the  result obtained from the Kraan-van Baal form of the solution. In Subsection \ref{app:A1} we remind the
reader the KvB formula and the equivalence with the our notation. Finally in Subsection \ref{app:A12} we explicit the computation
of the Taylor expansion. 
  
\subsection{The Kraan - van Baal (KvB) solution explicit}    
\label{app:A1}
 
The solution in terms of KvB prepotentials becomes, with $r_{12}=2\pi^2T^2\r^2$, $q_1=q,q_2=1-q$ and  rescaled coordinates \cite{Kraan:1998kp,Kraan:1998pm}:
 
 \ba
 \hat\psi&=&\frac{\sinh(q_1r_1)\sinh(q_2r_2)}{r_1r_2}\frac{1}{2}\left(\frac{2r_1r_2(\cosh (q_1 r_1)\cosh( q_2 r_2) -\cos t)}{\sinh (q_1r_1)\sinh (q_2r_2)} + r_1^2+r_2^2-r_{12}^2\right)\non\\
 \psi&=&\hat\psi+\frac{\sinh (q_1r_1 )\sinh (q_2r_2)}{r_1 r_2}r_{12}\left(r_{12} +r_1\cosh (q_1 r_1 )+  r_2\cosh (q_2 r_2 ) 
 \right)\non\\
 \phi&=& \frac{\psi}{\hat\psi} \,\,\,\, ~\mbox{(reduces to HS prepotential $\Pi$ in the $q_1\rightarrow 0$ limit.)} 
 \la{kvb1}
 \ea
 There is also a second -complex- prepotential:
 \ba
 \chi &=&{r_{12}\over{\psi}}\frac{\sinh (q_1r_1 )\sinh (q_2r_2)}{r_1 r_2}\left(e^{i q_1 t}\frac{r_1}{\sinh (q_1r_1)}+e^{-i q_2t}\frac{r_2}{\sinh (q_2r_2)}\right)
 \la{kvb2}
 \ea 
 The vector potential in terms of these prepotentials reads ($\t^\pm={1\over 2}(\t^1\pm i \t^2)$):
 \be
 A_\m^a{\t^a\over 2}=-\bar\eta^3_{\m\n}\t^3\partial_\n\log\phi -\phi(\t^+\bar\eta^-_{\m\n}\partial_\n\chi+\t^-\bar\eta_{\m\n}^+\partial_\n\chi^*).
 \ee
 Rotational invariance is gone and only rotations around the z-axis leave the caloron dipole configuration invariant.
 In particular the scalar potential equals
 \be
 A_0^a=\bar\eta^3_{0\n}\partial_\n\log\phi +\phi(\t^+\partial_-\chi+\t^-\partial_+\chi^*)\,.
 \ee
 We wrote $\partial_\pm=\partial_1\pm i\partial_2$
 
 The solution is invariant under  $q\leftrightarrow 1-q$ and moreover $ t \leftrightarrow -t $ for $\chi$.
     ~So the contribution to the free energy is Z(2) invariant.  
     
It is useful to rewrite the potential in terms of new prepotentials:
\begin{eqnarray}
 A_\mu = -\bar{\eta}^a_{\mu\nu}\hat F\mathcal D_\nu\left(\delta_{ab}
 V_{S} 
 +\epsilon_{a3b}
   V_{A} 
 +\delta_{3a}\delta_{3b}
  V_{3}\right)\frac{\tau^b}{2}
\end{eqnarray}
When computing the variation in $q$ it turns out that $\hat\psi$ starts out with a quadratic term in $q$. 
The other prepotentials  have  a complicated linear term,
but the result in terms of $V_S$, $V_A$, $V_3$ is relatively simple. It is given at the end of next subsection, Eq. (\ref{resulttaylorexpansion}).

\subsection{An insightful derivation of the  variation } 
\label{app:A12}  
In the notation introduced in \cite{GarciaPerez:2006rt,GarciaPerez:2009mg}. The SU(2) caloron gauge field is written just
in terms of a $2\times2$ real matrix, V, which inverse is additive in constituents monopoles:
\begin{eqnarray}
 A_\mu = \frac{i}{2}F
 \left(1 , i\tau_3\right)
 \sigma_\m\sigma_\n^\dagger\partial_\n V
 \left(\begin{array}{c}1 \\ -i\tau_3\end{array}\right) + \mathrm{h.c.}
\label{eq:GAGP}
 \end{eqnarray}
where \begin{eqnarray}   
V^{-1} &=& \frac{r_{12}}{4\pi^2}{\bf 1} +\sum_{a=1,2} \frac{r_a}{4\pi^2\sinh( m_a r_a)}
\left(\begin{array}{cc}
\cosh(q_ar_a) - \cos(q_a t) & (-1)^a \sin(q_a t) \\ 
 (-1)^a \sin(q_a t) & \cosh(q_ar_a) + \cos(q_a t)  
\end{array}\right) \nonumber
\end{eqnarray}   
and $F^{-1} = 1 - \frac{\rho^2}{2}\tr(V)$.  

Reordering Eq. \ref{eq:GAGP}, the vector potential can be written as a covariant derivative, up to a common factor:
\begin{eqnarray}  
 A_\mu = -\bar{\eta}^a_{\mu\nu}\hat F\mathcal D_\nu\left(\delta_{ab}
 V_{S}  +\epsilon_{a3b}   V_{A} +\delta_{3a}\delta_{3b}  V_{3}\right)\frac{\tau^b}{2}
\end{eqnarray}
where  
\begin{eqnarray}
\mathcal D_\mu  &=&  
 \partial_\mu - \partial_\mu \log \mathcal V,\,\,\, \mathcal V = \det V^{-1},\,\,\, \hat F = \frac{r_{12}}{4\pi^2}\frac{F}{\mathcal V}
  \end{eqnarray}
and
\begin{eqnarray}
V_S =   (V^{-1})_{11}- (V^{-1})_{22} &=&\sum_{a} \frac{r_a\cos(q_a t)}{2\pi^2\sinh(q_a r_a)} \label{eq:vs} \\
V_A = 2  (V^{-1})_{12} &=& -\sum_{a} \frac{(-1)^a r_a\sin(q_a t)}{2\pi^2\sinh(q_a r_a)}  \\
V_3 = 2  (V^{-1})_{11} &=& \frac{r_{12}}{2\pi^2} +  \sum_a \frac{r_a(\cosh(q_ar_a)- \cos(q_a t))}{2\pi^2\sinh(q_ar_a)}
\end{eqnarray}
where
\begin{eqnarray}
 q_1 &=& q,\,\,\,\, q_2 = 1 - q \\
 r_a^2 &=& |\vec x-\vec X_a|^2
 \end{eqnarray}
This notation is related to the  KvB  formululation by:
\begin{eqnarray}
 \hat\psi &=& K\frac{r_{12}}{\hat F} \\
 \psi &=& 4\pi^2 K \mathcal V \\
 \chi &=& \frac{ r_{12}}{\mathcal V}(V_S + i V_A) 
\end{eqnarray}
where $K = \frac{\sinh(q_1r_1)\sinh(q_2r_2)}{2r_1r_2}$.

The main advantage of this notation is that $\hat F$ is already order $q$, so we only need the leading order of the 
other quantities. We start computing the Taylor expansion of $\hat F$:        
\begin{eqnarray}
 \hat F = 2\pi^2q(\Pi-1)\left(
 1 - r_{12}q\left( \frac{r\coth(r)}{r_{12}}  + n_3 h  
 \right) \right) + \mathcal{O}(q^3)
\end{eqnarray}    
where $n_3 = \frac{x_3}{r}$ and $h$ is defined in Eq. \ref{eq:h}. Since $\hat F$ is already order $q$, we need also the order $q^2$ due to the $\frac{1}{q}$
term in Eq. \ref{eq:vs}.
In the case of $\mathcal V$ we only need the leading order:
\begin{eqnarray}
 \mathcal V = \frac{r_{12}}{4\pi^2}\left(\frac{1}{r_{12}q}\frac{\Pi}{\Pi-1}
 + 1 + \frac{r\coth( r)}{r_{12}}\frac{\Pi}{\Pi-1}  +  \frac{n_3 h}{\Pi-1} \right)
+ \mathcal{O}(q)
 \end{eqnarray}
 and for the derivative of logarithm we obtain:
\begin{eqnarray}
\partial_\mu\log \mathcal V = \partial_\mu\log\left(\frac{\Pi}{\Pi-1}\right)
 +r_{12}q\partial_\mu\left(\frac{\Pi-1}{\Pi}\left( 1 + \frac{r\coth(r)}{r_{12}}\frac{\Pi}{\Pi-1}  +  \frac{n_3h}{\Pi-1}\right)\right) \nonumber 
+ \mathcal{O}(q^2)
 \end{eqnarray}
Finally we can compute the leading order for the $V$  functions: 
\begin{eqnarray}
 V_S &=& \frac{1}{2\pi^2q}\left(1 - r_{12}q\left(\frac{1}{\Pi-1} - \frac{r\coth(r)}{r_{12}}\right)\right) + \mathcal{O}(q)\\
 V_A &=& \frac{t}{2\pi^2}  - \frac{r\sin( t)}{2\pi^2\sinh(r)} + \mathcal{O}(q) \equiv V_A^0 + \mathcal{O}(q)\\
 V_3 &=& \frac{r_{12}}{2\pi^2}\frac{\Pi}{\Pi-1} + \mathcal{O}(q)
\end{eqnarray}
Combining the previous expansion we arrive to the final results:
\begin{eqnarray}
\hat F\mathcal D_\mu V_S &=&\partial_\mu\left(\log \Pi - r_{12} q\frac{1}{\Pi}\right)
- r_{12}q\partial_\mu\left(n_3h \frac{(\Pi-1)}{\Pi} 
 \right) + \mathcal{O}(q^2) \\
\hat F \mathcal D_\mu V_A &=&  
 2\pi^2q\left( (\Pi-1)\partial_\mu V_A^0+  V_A^0 \partial_\mu  \log \Pi\right) + \mathcal{O}(q^2) \\
\hat  F \mathcal D_\mu V_3 &=& \mathcal{O}(q^2) 
\la{resulttaylorexpansion}
 \end{eqnarray}

\section{Details for the  analytic form of  the Harrington-Shepard determinant} \la{app:B}

In this appendix we resume the results obtained in \cite{Altes:2014bwa} for the computation of 
the determinant for caloron gauge fields that satisfy:
\begin{eqnarray}
\partial_\alpha A_\mu^a = -c_\alpha\bar\eta^a_{\mu\nu} \frac{\partial_\mu \Pi}{\Pi^2}
\end{eqnarray}   
We will be interested in the cases $\alpha= r_{12}$ and $\alpha = q$, with $c_\a=\frac{1}{r_{12}}$, $r_{12}$ respectively. 

The variation of the determinant for spin 1 is given by the formula \cite{Brown:1977cm,Brown:1977eb,Brown:1978yj,Gross:1980br,Altes:2014bwa}:
\begin{eqnarray}
 \delta_\a\log \det \left(-D^2\right) &=&
  \int_0^\alpha d\alpha' \left(4 A_{\alpha'} + B_{\alpha'} + c_{\alpha'}\frac{2}{3}r_{12}\right)  
\label{eq:vadet}
 \end{eqnarray} 
where  \cite{Altes:2014bwa}    
\begin{eqnarray}   
   A_\alpha &=&  
 \frac{c_\alpha}{12}\frac{1}{4\pi^2} \int_{R^3\times S^1} dx^4 \frac{\left((\partial_\mu \Pi)^2\right)^2}{\Pi^5} \\
  B_\alpha  &=&  
 \frac{c_\alpha}{4\pi^2}\int_{R^3\times S^1} dx^4 \frac{\left((\partial_\mu \Pi)^2\right)^2}{\Pi^5}H \\
 H &=& \frac{h^2-1}{h^2-1+2f}   \\    
 h &=& \coth( r) - \frac{1}{ r} \label{eq:h}
 \end{eqnarray}  
The integration in $r_{12}$ is easily done noting that:    
 \begin{eqnarray}         
 \frac{ (\partial_\mu \Pi)^2}{\Pi^2}= \frac{r_{12}^2f^2}{(1 + r_{12}f)^2}(h^2-1+2f)
 \end{eqnarray}
So we have:
\begin{eqnarray}  
\bar A(r_{12})=\int_0^{r_{12}} d r_{12}'  \frac{1}{r_{12}'} A_{r_{12}}(\lambda) &=& 
 \frac{1}{12}\frac{1}{16\pi^2}\frac{\left((\partial_\mu \Pi)^2\right)^2}{\Pi^4} \label{eq:A}\\
B(r_{12}) = \int_0^{r_{12}} d r_{12}' \frac{1}{r_{12}'}B_{r_{12}}(\lambda) &=& \frac{1}{16\pi^2}
\frac{\left((\partial_\mu \Pi)^2\right)^2}{\Pi^4}H(r,t) \label{eq:b}
\end{eqnarray}        
$\bar A(r_{12})$ contains a logarithmic singularity and can be regularized using the single zero-temperature instanton \cite{Brown:1978yj,Gross:1980br}
\begin{eqnarray}
{A}(r_{12}) &=& 
 \frac{1}{12}\frac{1}{16\pi^2}\left(\int_{R^3\times S^1} dx^4\frac{\left((\partial_\mu \Pi)^2\right)^2}{\Pi^4} - \int_{R^4} dx^4
  \frac{\left((\partial_\mu \Pi_0)^2\right)^2}{\Pi_0^4}\right)
\end{eqnarray}        
Numerically, it has been observed \cite{Gross:1980br,Altes:2014bwa} that the relation $B(r_{12})=12 A(r_{12})$ holds, although there is not 
a formal proof of this relation yet.
$A(r_{12})$ was parametrized in very good agreement through the functional forms \cite{Gross:1980br}:    
\begin{eqnarray}    
 A(\lambda) &=& -\frac{1}{12}\log\left(1 + \frac{r_{12}}{6}\right) + 0.0129\left(1 + 0.899\left(\frac{r_{12}}{2}\right)^{-3/4}\right)^{-8}\,.   
\end{eqnarray}       
 The   logarithmic terms  dominate over the whole range.

For the variation in terms of the holonomy $(c_\alpha = r_{12})$, we obtain
\begin{eqnarray}
 \delta_q\log \det \left(-D^2\right)|_{q=0} &=&
  qr_{12} \left(4 \tilde A + \tilde B + \frac{2}{3}r_{12}\right)  
 \end{eqnarray} 
 where  
\begin{eqnarray}     
 \tilde A &=&  
 \frac{1}{12}\frac{1}{4\pi^2} \int_{R^3\times S^1} dx^4 \frac{\left((\partial_\mu \Pi)^2\right)^2}{\Pi^5} \label{eq:ta} \\
 \tilde B  &=&       
 \frac{1}{4\pi^2}\int_{R^3\times S^1} dx^4 \frac{\left((\partial_\mu \Pi)^2\right)^2}{\Pi^5}H  \label{eq:tb}
\end{eqnarray}       
We find a numerical relation $\tilde{B} = 12 \tilde A -1$. Combining Eqs. (\ref{eq:b}), (\ref{eq:barA}), (\ref{eq:ta}) and (\ref{eq:tb}) and using 
$\tilde B = r_{12}\partial_{r_{12}} B$, we obtain our final result for the determinants:   
\begin{eqnarray}
 \delta_{r_{12}}\log \det \left(-D^2_{1}\right) &=& 
   16 A(r_{12}) + \frac{2}{3}r_{12}\,,   \\
 \delta_q\log \det \left(-D^2\right)|_{q=0} &=& 
  q r_{12}^2\partial_{r_{12}}  \left(16  A(r_{12}) +  \frac{1}{3}\log(r_{12})   + \frac{2}{3}r_{12}\right)  
\end{eqnarray}

  \bibliographystyle{JHEP}
  \bibliography{mskvbvar}
\end{document}